\documentclass[pdflatex,sn-mathphys-num]{sn-jnl}

\usepackage{graphicx}%
\usepackage{multirow}%
\usepackage{amsmath,amssymb,amsfonts}%
\usepackage{amsthm}%
\usepackage[title]{appendix}%
\usepackage{xcolor}%
\usepackage{textcomp}%
\usepackage{manyfoot}%
\usepackage{booktabs}%
\usepackage{algorithm}%
\usepackage{algorithmicx}%
\usepackage{algpseudocode}%
\usepackage{listings}%
\usepackage{array}%
\usepackage{url}%
\newcolumntype{L}[1]{>{\raggedright\arraybackslash}p{#1}}%
\newcolumntype{C}[1]{>{\centering\arraybackslash}p{#1}}%
\newcommand{\pmc}{\!\pm\!}%
\newcommand{\best}[1]{{\bfseries\boldmath #1}}%
\newcommand{\second}[1]{\underline{{\bfseries\boldmath #1}}}%

\theoremstyle{thmstyleone}%
\theoremstyle{thmstyletwo}%

\theoremstyle{thmstylethree}%

\begin{document}

\title[RankGLU]{RankGLU: Residual Gated Score Formation for Cross-Sectional Stock Prediction}


\author*[1]{\fnm{Huixiang} \sur{Xiao}}\email{Huixiangxiao.cs@gmail.com}

\author[2]{\fnm{Jian} \sur{ Xu}}\email{jian.xu@riken.jp}

\author[3]{\fnm{Feiyu} \sur{Qu}}\email{202300450070@mail.sdu.edu.cn}

\author[4]{\fnm{Zixuan} \sur{Xie}}\email{zixuanxie90@gmail.com}

\author[5]{\fnm{Xiangyu} \sur{Li}}\email{xiangyuli@sjtu.edu.cn}

\affil*[1]{\orgname{College of Computer Science and Engineering,Chongqing University of Technology}, \orgaddress{\city{Chongqing}, \postcode{402160}, \country{China}}}

\affil[2]{\orgname{RIKEN}, \orgaddress{\city{Wako}, \postcode{3510198}, \country{Japan}}}

\affil[3]{\orgname{School of Control Science and Engineering, Shandong University}, \orgaddress{\city{Jinan}, \postcode{250061}, \country{China}}}

\affil[4]{\orgname{School of Communication and Information Engineering,Chongqing Yitong College}, \orgaddress{\street{Hechuan Campus}, \city{Chongqing}, \postcode{401520}, \country{China}}}

\affil[5]{\orgdiv{Department of Electronic Engineering}, \orgname{Shanghai Jiao Tong University}, \orgaddress{\city{Shanghai}, \postcode{200240}, \country{China}}}


\abstract{Cross-sectional stock prediction is closer to a ranking problem than to ordinary return-magnitude regression, since portfolio decisions depend on the relative ordering of assets within each trading date. Existing temporal, graph-based, and market-conditioned attention models have improved stock representation learning, yet the final prediction head is often treated as a minor implementation detail. This paper argues that, under information-coefficient-oriented evaluation, score formation is a critical bottleneck: an over-flexible head can fit unstable return magnitude, whereas an overly linear head may underuse cross-feature interactions. We therefore develop RankGLU, a residual bottleneck gated linear unit for cross-sectional stock ranking. RankGLU keeps a direct linear scoring path and adds a bounded multiplicative branch, thereby preserving a stable ordering route while allowing controlled nonlinear interactions. The method is evaluated on CSI300 and CSI800 under a unified protocol with cross-sectional score normalization and an IC-augmented objective. Multi-seed experiments show that, on CSI300, RankGLU achieves the strongest mean IC among the internally controlled variants, improving from 0.0654$\pm$0.0052 for the original backbone and 0.0697$\pm$0.0030 for the ranking-aware backbone to 0.0727$\pm$0.0037, a gain that is consistent across all five seeds. Its best-seed result also exceeds the corresponding baselines. Ablation results further indicate that removing the GLU prediction head causes the clearest degradation among the tested component changes. Additional relation-path calibrations can produce high single-seed peaks, but their multi-seed behavior is less stable. The evidence suggests that ranking-aware stock models benefit most reliably from bounded residual score formation rather than from indiscriminate architectural expansion. Code will be released at \url{https://github.com/HuixiangXiao/RankGLU}.}




\maketitle

\section{Introduction}\label{sec1}

Modern financial information systems often need to convert heterogeneous market observations into cross-sectional ranking signals. In stock prediction, the practical decision is rarely whether the exact next-period return of a single asset can be regressed with high numerical precision. Portfolio construction usually requires a model to identify, on each trading day, which assets are relatively stronger or weaker than their peers, which is consistent with the cross-sectional view of expected return predictability in empirical asset pricing \cite{fama1992cross,fama1993common,jegadeesh1993returns,carhart1997persistence}. The task is therefore naturally related to noisy cross-sectional ranking as well as point-wise return regression. Accordingly, widely used evaluation criteria in quantitative investment, such as the information coefficient (IC), rank information coefficient (RankIC), IC information ratio (ICIR), and long-short portfolio performance, emphasize relative ordering rather than absolute return magnitude.

This distinction is relevant because financial returns are low-signal, heavy-tailed, and regime-dependent. The absolute magnitude of realized returns is often entangled with volatility, liquidity, market-wide shocks, idiosyncratic events, and temporary scale drift; related evidence from return-predictability and machine-learning asset-pricing studies also suggests that apparent predictive signals can be fragile under repeated testing and changing market regimes \cite{mclean2016does,harvey2016cross,gu2020empirical,chen2026esgalpha}. A model optimized mainly to fit return magnitude may therefore allocate part of its capacity to unstable noise. This mismatch can appear in different parts of a deep stock model, but it becomes especially consequential at the final score-formation stage. After temporal and cross-stock encoders have produced a stock embedding, the prediction head determines how that embedding is converted into the ranking score used by IC and RankIC. If the head is too weak, useful cross-feature interactions may be underused; if it is too flexible, unstable magnitude patterns can be mapped into noisy scores.

Existing relation-aware stock forecasting studies can be broadly divided into graph-based, hypergraph-based, attention-based, and market-conditioned Transformer-style models. These methods suggest that stock movements should not be modeled independently and that useful relations can be dynamic rather than fixed by industry or other static concepts. A recent market-conditioned temporal--relational architecture also indicates that market states can guide feature selection and that alternating intra-stock and inter-stock aggregation is a practical way to model momentary and cross-time stock relations. However, most of these studies primarily discuss how representations are extracted, while the final score head is often treated as a small implementation component. This leaves a methodological gap between \emph{representation learning} and \emph{ranking-aware score formation}.

RankGLU is designed for this gap without introducing an additional industry branch, market branch, or external relation module. We retain the market-conditioned temporal--relational decomposition, but focus on the final operation that maps the stock embedding to a cross-sectional score. The design principle is conservative: the prediction head should preserve a direct scoring path for stable ordering while adding only bounded nonlinear interactions. Concretely, RankGLU replaces the enhanced MLP-style prediction head with a residual bottleneck GLU head,
\begin{equation}
\hat{y}_i =
W_s \mathrm{LN}(h_i)
+\gamma W_o\left[
\left(W_v\mathrm{LN}(h_i)\right)
\odot
\sigma\left(W_g\mathrm{LN}(h_i)\right)
\right],
\label{eq:intro_rankglu_head}
\end{equation}
where the first term is a direct residual scoring path and the second term is a bounded gated interaction path. The bottleneck controls the capacity of the nonlinear branch, and $\gamma$ controls its strength. This construction is aligned with the IC-oriented setting: the model can express feature interactions, but it is not forced to route all scores through an unconstrained nonlinear decoder.

The empirical analysis is organized in a way that separates stable improvement from exploratory peak performance. First, the original reproduced backbone provides the architectural reference. Second, a ranking-aware backbone introduces cross-sectional score normalization and an IC-augmented objective, aligning the training protocol with the evaluation metric. Third, RankGLU is evaluated as the retained component replacement on top of that ranking-aware backbone. Under five random seeds on CSI300, RankGLU obtains the highest mean IC among the internally controlled variants, improving from $0.0654\pm0.0052$ for the original backbone and $0.0697\pm0.0030$ for the ranking-aware backbone to $0.0727\pm0.0037$. The best-seed result of RankGLU also remains above the corresponding baseline peaks. In contrast, several relation-path calibrations, including cosine inter-stock scores and centered value gates, can reach higher single-seed peaks but show less favorable multi-seed averages. We therefore report these relation-path modifications as diagnostic explorations rather than as the main retained method.

\begin{figure*}[t]
\centering
\includegraphics[width=0.96\textwidth]{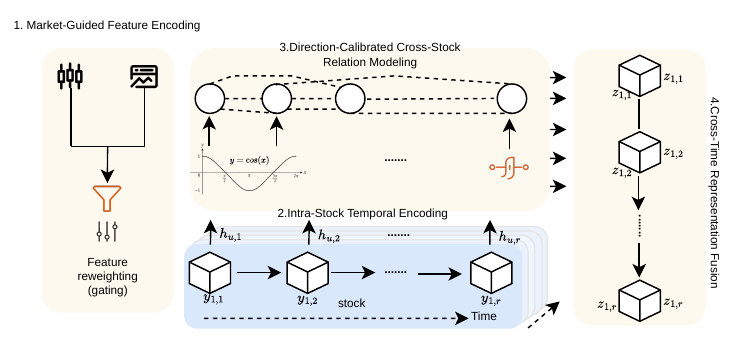}
\caption{Overview of the studied temporal--relational ranking framework. The retained RankGLU design keeps the market-guided and temporal--relational backbone and applies residual GLU-based score formation at the prediction head. The inter-stock relation calibrations shown in the cross-stock block are treated as diagnostic extensions in the experiments.}
\label{fig:rankglu_overview}
\end{figure*}

Figure~\ref{fig:rankglu_overview} summarizes the computation flow and the explored component locations. The main retained design keeps the four-stage information path from market-guided feature encoding to intra-stock temporal encoding, inter-stock relation aggregation, and final cross-time aggregation with score formation. The decisive replacement is placed at the prediction stage, where the residual GLU head forms the final ranking score through a stable linear branch and a bounded gated-interaction branch. This view is also consistent with several diagnostic observations. Sparse attention normalizers such as entmax or sparsemax may remove too many cross-stock relations in a market where weak signals are distributed. Extra relation-path calibration can produce strong single-seed values but is more sensitive to random initialization. Feature-layer and temporal gates may disturb representations before the final ranking geometry is formed. By contrast, the prediction head is directly coupled to the score vector evaluated by IC and RankIC, making it a more reliable location for component-level improvement.

The contributions of this work are summarized as follows. First, we formulate a ranking-aware score-formation perspective for low signal-to-noise cross-sectional stock prediction. Second, using the enhanced temporal--relational baseline as a controlled starting point, we propose RankGLU, a residual bottleneck GLU prediction head that combines a direct linear scoring path with bounded multiplicative interactions. Third, we provide multi-seed evidence showing that RankGLU delivers the strongest CSI300 mean IC among the internally controlled variants and retains favorable best-seed behavior. Finally, diagnostic and ablation results are used to distinguish robust prediction-head improvement from less stable relation-path and sparsification alternatives.

\section{Related Work}\label{sec:related_work}

This section reviews the literature most closely related to RankGLU. The discussion is organized around four aspects: deep sequential models for stock prediction, relation-aware stock forecasting, ranking-oriented financial evaluation, and the calibration of attention and gated prediction operators.

\subsection{Deep learning for stock prediction}\label{subsec:rw_deep_stock}

Deep neural networks have been increasingly used to model financial time series because they provide flexible nonlinear mappings from historical observations to future return proxies. Earlier machine-learning stock prediction studies used support vector machines and other nonlinear classifiers to predict market direction and stock movement \cite{tay2001svm,kim2003svm,kara2011direction,patel2015predicting}. Deep representation learning later became attractive because it can extract latent temporal patterns from raw or engineered market features \cite{chong2017deep,krauss2017deep,fischer2018lstm,zhang2019deeplob,sezer2020systematic}. Recurrent architectures such as long short-term memory networks and gated recurrent units were originally designed to alleviate the vanishing-gradient problem in sequential modeling \cite{hochreiter1997lstm,cho2014gru}. They have therefore become common choices for stock movement prediction, where temporal dependence, momentum reversal, and short-term persistence may all be informative. Convolutional sequence models, including temporal convolutional networks, provide another route by extracting local temporal patterns through causal convolution and dilation \cite{bai2018tcn}. More recently, Transformer-style self-attention has been adopted because it can model long-range dependencies without recurrent state propagation \cite{vaswani2017attention}. Recent financial-language and time-series representation studies further indicate that external information and sequence-to-language abstractions can complement purely numerical features \cite{huang2026finsent,xun2026ts2lang}.

These models, however, mainly focus on the temporal evolution of each asset. When they are applied independently or with weak cross-sectional coupling, they tend to treat stocks as separate sequences and leave the relation among assets under-modeled. This limitation is relevant in equity markets because price movements are often affected by market-wide shocks, sector-level co-movement, supply-chain relations, liquidity spillovers, and other cross-stock interactions. A purely temporal model may therefore capture part of the within-stock signal while missing some of the relative information needed for daily cross-sectional ranking.

\subsection{Relation-aware and graph-based stock forecasting}\label{subsec:rw_relation_graph}

To incorporate cross-stock information, a growing body of work has considered relation-aware and graph-based stock forecasting. Graph neural networks and graph attention networks provide a direct mechanism for aggregating information from neighboring stocks \cite{velickovic2018graph}. In financial applications, the graph can be constructed from industry membership, supply-chain relations, shareholder relations, price co-movement, or dynamically estimated correlations. Earlier relation-ranking models have reported that explicitly using stock relations can improve the prediction of relative returns \cite{feng2019temporal}. Hypergraph-based methods further generalize pairwise edges to higher-order relations, allowing a group of stocks to share latent market or sector factors \cite{sawhney2020hypergraph}. Hybrid sequential and graph-based designs have also been explored to jointly capture temporal and relational dependencies \cite{mao2025hybrid}.

While these models can improve upon isolated sequence modeling, they also raise two practical issues. First, fixed relation graphs may be incomplete or stale because stock relations evolve with market regimes. Second, aggressive graph sparsification can remove weak but collectively informative relations. Recent temporal--relational attention models address part of this problem by alternating intra-stock and inter-stock aggregation and by using market information to guide feature selection \cite{li2024master}. RankGLU follows this general decomposition but focuses on a different bottleneck: even when the temporal--relational encoder is structurally appropriate, the final mapping from representation to ranking score may still be poorly aligned with IC-oriented evaluation. This observation motivates the prediction-head-centered design developed in this paper.

\subsection{Ranking-oriented financial prediction}\label{subsec:rw_ranking}

Financial forecasting differs from ordinary regression because the final decision is typically a cross-sectional ranking or portfolio allocation. The empirical asset-pricing literature has long emphasized that expected returns are identified through cross-sectional differences rather than through the exact point prediction of each realized return \cite{fama1992cross,jegadeesh1993returns,carhart1997persistence,kelly2019characteristics}. Metrics such as IC, RankIC, ICIR, and RankICIR are designed to measure whether predicted scores preserve the relative ordering of future returns across stocks. Portfolio-oriented metrics such as annualized return and information ratio further evaluate whether the ranking can be converted into practical long-short performance. From this perspective, an accurate estimate of absolute return magnitude is less important than a stable ordering of assets within each trading-day cross-section.

This distinction has methodological consequences. If a model is trained only with point-wise mean squared error, high-magnitude return observations can dominate the loss even when they are not representative of stable alpha. Modern machine-learning asset-pricing studies also suggest that regularization and representation constraints are important when predictive signals are numerous but weak \cite{kozak2020shrinking,gu2020empirical,gu2021autoencoder}. Cross-sectional normalization and correlation-oriented objectives are therefore used to align training with ranking evaluation. In the enhanced baseline protocol used in this study, this idea is operationalized by combining cross-sectional score normalization with an IC-augmented objective. The design does not introduce an additional prediction target; rather, it aligns the optimization geometry with the evaluation rule that is already used in quantitative stock selection.

\subsection{Attention calibration and gated prediction heads}\label{subsec:rw_attention_gate}

Attention mechanisms are useful because they adaptively select relevant context. Standard dot-product attention computes a score by $q^\top k$, which can be decomposed into the product of query norm, key norm, and directional similarity. In many natural language or vision tasks, the norm term may encode useful confidence or salience. In stock prediction, however, representation norms may also reflect volatility shocks, liquidity disturbances, or day-level scale drift. Cosine-based attention removes this direct norm dependence from the score path and therefore provides a more direction-oriented relation measure.

Gated neural components provide another way to regulate information flow. The gated linear unit introduces multiplicative interactions through a sigmoid gate \cite{dauphin2017glu}, while later variants such as SwiGLU and gated MLPs have been used to increase the expressiveness of feed-forward blocks \cite{shazeer2020glu}. For financial prediction, the relevant question is less whether a stronger gate is universally more expressive than whether the gate provides a stable inductive bias under noisy labels. RankGLU is designed from this perspective. It keeps a direct linear scoring path and adds only bounded multiplicative interactions through a bottleneck GLU branch. This makes the final score head expressive enough to model feature interactions while reducing the risk that unstable return magnitude dominates the ranking score.

\subsection{Relevance to intelligent financial information systems}\label{subsec:rw_intelligent_systems}

From the perspective of intelligent information systems, cross-sectional stock prediction is not only a forecasting problem but also a ranking, retrieval, and decision-support problem over a continuously updated financial database. A practical stock-selection system must integrate heterogeneous market features, market-state information, relational dependencies among assets, and uncertainty-aware ranking scores into a coherent information-processing pipeline. This requirement is closely aligned with research on intelligent information systems, where machine learning components are expected to support robust information filtering, adaptive decision making, and reliable retrieval from noisy and dynamic data sources.

RankGLU contributes to this systems-oriented setting by focusing on the final score-formation interface between representation learning and downstream ranking decisions. Instead of treating the neural prediction head as an isolated numerical regressor, the proposed design calibrates how latent financial information is converted into a cross-sectional score vector. The resulting score can be directly consumed by portfolio construction, risk screening, or human-in-the-loop investment decision-support modules. This positioning also clarifies why the method emphasizes multi-seed stability and component-level attribution: for an intelligent financial information system, a deployable ranking module should be not only expressive but also robust under repeated training and changing market conditions.

\section{Methodology}\label{sec:methodology}

This section formalizes RankGLU. The method is deliberately placed at the final score-formation stage rather than as an additional side branch. We first define the cross-sectional ranking problem and the market-conditioned temporal--relational backbone used as the representation extractor. We then describe the ranking-aware training protocol and the residual bottleneck GLU prediction head that constitutes the retained component replacement.

\subsection{Problem Formulation}\label{subsec:problem_formulation}

Let $\mathcal{S}_{d}$ denote the tradable stock universe on prediction date $d$, and let $N_d=|\mathcal{S}_{d}|$. For each stock $u\in\mathcal{S}_{d}$, the model observes a historical feature window
\begin{equation}
X_{u,d}=\left[x_{u,d-\tau+1},x_{u,d-\tau+2},\ldots,x_{u,d}\right]\in\mathbb{R}^{\tau\times F},
\label{eq:problem_feature_window}
\end{equation}
where $\tau$ is the lookback length and $F$ is the number of stock-level features. The market state vector associated with date $d$ is denoted by $m_d\in\mathbb{R}^{F_m}$. The future return label over the prediction horizon is denoted by $r_{u,d+h}$, or simply $r_{u,d}$ when the horizon is clear from context. The learning problem is to estimate a scalar score
\begin{equation}
\hat r_{u,d}=f_{\theta}(X_{u,d},m_d,\mathcal{S}_{d}),
\label{eq:problem_score}
\end{equation}
whose cross-sectional ordering is aligned with the future labels $\{r_{u,d}\}_{u\in\mathcal{S}_{d}}$.

Unlike ordinary regression, the primary object is not the exact magnitude of $\hat r_{u,d}$ but the ranking induced by the vector
\begin{equation}
\hat{\boldsymbol r}_{d}=\{\hat r_{u,d}:u\in\mathcal{S}_{d}\}.
\label{eq:problem_score_vector}
\end{equation}
For this reason, the information coefficient on date $d$ is defined as
\begin{equation}
\operatorname{IC}_{d}
=
\operatorname{corr}\left(\hat{\boldsymbol r}_{d},\boldsymbol r_{d}\right),
\label{eq:problem_ic}
\end{equation}
and the overall objective is to increase the average cross-sectional correlation across testing dates. This formulation makes explicit the main mismatch considered in RankGLU: hidden representations may contain useful magnitude information, but the final evaluation is largely determined by relative cross-sectional ordering.

\subsection{Market-Conditioned Temporal--Relational Backbone}\label{subsec:backbone}

The prior market-conditioned temporal--relational architecture contains five successive stages: market-conditioned feature gating, intra-stock aggregation, inter-stock aggregation, temporal aggregation, and prediction. We retain this high-level decomposition because it is well matched to stock forecasting: market states first regulate feature salience, intra-stock aggregation extracts temporal patterns within each asset, inter-stock aggregation transfers information from momentarily correlated assets, temporal aggregation pools the resulting sequence, and the prediction head maps the final embedding to a stock score.

Following the original market-conditioned feature gate, the market state is transformed into a feature-wise scaling vector:
\begin{equation}
\alpha(m_{\tau}) = F \cdot \operatorname{softmax}\left(\frac{W_{\alpha}m_{\tau}+b_{\alpha}}{\beta}\right),
\label{eq:market_gate}
\end{equation}
where $\beta$ controls the sharpness of feature selection. The feature tensor is then rescaled as
\begin{equation}
\tilde{x}_{u,t} = \alpha(m_{\tau}) \odot x_{u,t}.
\label{eq:market_rescale}
\end{equation}
This gate is preserved in RankGLU because it provides a market-conditioned feature selection prior. Diagnostic experiments with sparse normalizers in this gate suggest that overly sparse feature selection can discard distributed weak factors.

The rescaled features are projected into a $D$-dimensional hidden space and encoded by intra-stock attention:
\begin{equation}
y_{u,t}=f(\tilde{x}_{u,t})+p_t,\quad
H^1_u=\operatorname{FFN}_1\left(\operatorname{MHA}_1(Y_u)+Y_u\right),
\label{eq:intra_stock}
\end{equation}
where $p_t$ is a sinusoidal positional encoding and $Y_u=\{y_{u,t}\}_{t=1}^{\tau}$.

At each time step, the hidden states of all stocks are then aggregated by the inter-stock attention block. Let $h_{u,t}$ denote the temporal embedding of stock $u$ at time $t$. The relation logit and attention weight are
\begin{equation}
\ell_{uv,t}
=
\frac{q_{u,t}^{\top}k_{v,t}}{\sqrt{d_h}},
\quad
a_{uv,t}
=
\frac{\exp(\ell_{uv,t})}
{\sum_{v'\in\mathcal{S}_{d}}\exp(\ell_{uv',t})},
\label{eq:inter_stock_original}
\end{equation}
where $q_{u,t}$ and $k_{v,t}$ are query and key vectors. The propagated message is added through the original residual path:
\begin{equation}
z_{u,t}
=
\operatorname{LN}
\left(
h_{u,t}+\sum_{v\in\mathcal{S}_{d}}a_{uv,t}W_Vh_{v,t}
\right),
\quad
z_{u,t}\leftarrow z_{u,t}+\operatorname{FFN}_{2}(z_{u,t}).
\label{eq:inter_stock_output_original}
\end{equation}
The sequence $\{z_{u,t}\}_{t=1}^{\tau}$ is summarized by temporal aggregation to obtain the stock-level embedding $e_u$. In the retained RankGLU model, these representation stages are not enlarged. This choice isolates the main architectural difference at the score head.

\subsection{Ranking-Aware Objective and Cross-Sectional Normalization}\label{subsec:rank_objective}

The ranking-aware baseline used in this study applies an IC-augmented objective and cross-sectional output normalization. This setting is part of the reference protocol on which RankGLU is built, rather than the prediction-head replacement itself. For each trading date, predictions and labels form a cross-section. Given valid predictions $\hat{\boldsymbol r}$ and labels $\boldsymbol r$, the Pearson correlation used by IC is
\begin{equation}
\rho(\hat{\boldsymbol r},\boldsymbol r)
=
\frac{(\hat{\boldsymbol r}-\bar{\hat r}\mathbf{1})^\top(\boldsymbol r-\bar r\mathbf{1})}
{\|\hat{\boldsymbol r}-\bar{\hat r}\mathbf{1}\|_2\|\boldsymbol r-\bar r\mathbf{1}\|_2}.
\label{eq:ic_corr}
\end{equation}
The optimized loss is
\begin{equation}
\mathcal{L}
=
\operatorname{MSE}(\hat{\boldsymbol r},\boldsymbol r)
\,+\,
\lambda\left(1-\rho(\hat{\boldsymbol r},\boldsymbol r)\right),
\label{eq:mse_ic_loss}
\end{equation}
where $\lambda=0.1$ in all reported experiments. In addition, the model output is normalized within each daily cross-section:
\begin{equation}
\operatorname{CSNorm}(\hat{\boldsymbol r})
=
\frac{\hat{\boldsymbol r}-\mu(\hat{\boldsymbol r})}
{\sigma(\hat{\boldsymbol r})+\epsilon}.
\label{eq:cs_norm}
\end{equation}
If both prediction and label are cross-sectionally standardized, the squared error satisfies
\begin{equation}
\operatorname{MSE}(\hat{\boldsymbol r},\boldsymbol r)
=
\frac{1}{N}\sum_{u=1}^{N}(\hat r_u-r_u)^2
=2\left(1-\rho(\hat{\boldsymbol r},\boldsymbol r)\right),
\label{eq:mse_ic_relation}
\end{equation}
up to numerical constants. Under this normalization, the training target becomes closely aligned with IC-based ranking evaluation. This observation motivates a prediction head that preserves a stable scoring geometry rather than an unrestricted magnitude-fitting decoder.

\subsection{RankGLU Prediction Head}\label{subsec:glu_head}

The prediction module is the location where representation magnitude is finally converted into a tradable ranking score. The original backbone uses a linear predictor, while the ranking-aware baseline uses an MLP-style nonlinear head. RankGLU replaces this final mapping with a residual bottleneck GLU module:
\begin{equation}
\tilde{e}_u=\operatorname{LN}(e_u),
\quad
\hat r_u
=
w_s^{\top}\tilde{e}_u
\,+\,
\gamma\,w_o^{\top}
\left[
(W_v\tilde{e}_u)\odot\sigma(W_g\tilde{e}_u)
\right],
\label{eq:residual_glu_head_method}
\end{equation}
where $e_u$ is the temporally pooled stock embedding, $\gamma$ is the GLU scale, and the hidden dimension of $W_v$ and $W_g$ is set by the decoder bottleneck. In the main configuration, the bottleneck is set to 128 according to the multi-seed comparison. The first term keeps a direct linear scoring path, while the second term introduces bounded multiplicative interactions. A local expansion of the sigmoid gate,
\begin{equation}
\sigma(W_g\tilde{e}_u)
\approx
\frac{1}{2}+\frac{1}{4}W_g\tilde{e}_u,
\label{eq:glu_expansion}
\end{equation}
gives
\begin{equation}
(W_v\tilde{e}_u)\odot\sigma(W_g\tilde{e}_u)
\approx
\frac{1}{2}W_v\tilde{e}_u
+\frac{1}{4}
\left(W_v\tilde{e}_u\right)\odot
\left(W_g\tilde{e}_u\right).
\label{eq:rankglu_second_order}
\end{equation}
Thus the gated branch can be interpreted as a low-rank mixture of a linear correction and a bounded second-order interaction. This property is useful for financial prediction because it increases expressiveness without forcing all scores through an unconstrained nonlinear decoder or an unbounded gate such as SwiGLU.

The residual form also gives a useful correlation-level interpretation. Let
\begin{equation}
a_u=w_s^{\top}\tilde{e}_u,\qquad
b_u=w_o^{\top}\left[(W_v\tilde{e}_u)\odot\sigma(W_g\tilde{e}_u)\right],
\label{eq:head_decomp}
\end{equation}
so that $\hat r_u=a_u+\gamma b_u$. For one trading-day cross-section, denote the centered vectors by $\boldsymbol a$, $\boldsymbol b$, and $\boldsymbol r$. The IC of the final score can be written as
\begin{equation}
\rho(\boldsymbol a+\gamma\boldsymbol b,\boldsymbol r)
=
\frac{
\operatorname{cov}(\boldsymbol a,\boldsymbol r)
+\gamma\operatorname{cov}(\boldsymbol b,\boldsymbol r)}
{\sigma_{\boldsymbol r}
\sqrt{
\sigma_{\boldsymbol a}^{2}
+2\gamma\operatorname{cov}(\boldsymbol a,\boldsymbol b)
+\gamma^{2}\sigma_{\boldsymbol b}^{2}}}.
\label{eq:rankglu_ic_decomp}
\end{equation}
Equation~\eqref{eq:rankglu_ic_decomp} shows why the nonlinear branch should be residual rather than fully replacing the score path. The branch improves IC only when it contributes label-aligned covariance without introducing excessive variance or collinearity with the linear path. Around $\gamma=0$, the first-order sensitivity is
\begin{equation}
\left.
\frac{\partial \rho(\boldsymbol a+\gamma\boldsymbol b,\boldsymbol r)}
{\partial \gamma}
\right|_{\gamma=0}
=
\frac{
\operatorname{cov}(\boldsymbol b,\boldsymbol r)\sigma_{\boldsymbol a}^{2}
-\operatorname{cov}(\boldsymbol a,\boldsymbol r)\operatorname{cov}(\boldsymbol a,\boldsymbol b)}
{\sigma_{\boldsymbol r}\sigma_{\boldsymbol a}^{3}}.
\label{eq:rankglu_ic_gradient}
\end{equation}
This expression makes the design requirement explicit: the gated branch should add a complementary ranking signal, not merely amplify an already noisy magnitude direction. The bottleneck dimension and the sigmoid gate are therefore used as structural controls on the covariance contribution of $\boldsymbol b$.

\subsection{Overall Pipeline}\label{subsec:overall_pipeline}

The retained RankGLU model can be summarized as
\begin{equation}
\hat{\boldsymbol r}
=
\operatorname{CSNorm}
\left[
g_{\mathrm{RankGLU}}
\left(
\operatorname{TempAttn}
\left(
\operatorname{InterAttn}
\left(
\operatorname{TAttn}
\left(f(\alpha(m_{\tau})\odot X)\right)
\right)
\right)
\right)
\right],
\label{eq:overall_pipeline}
\end{equation}
where $\operatorname{InterAttn}$ denotes the retained inter-stock aggregation block and $g_{\mathrm{RankGLU}}$ denotes the residual bottleneck GLU prediction head. The design can be viewed as a targeted score-head replacement on top of a market-conditioned temporal--relational paradigm, rather than as an external side branch or a broad architectural expansion.

\subsection{Algorithm Procedure}\label{subsec:algorithm_procedure}

Algorithm~\ref{alg:rankglu} summarizes the full training procedure. The algorithm is written at the level of trading-day grouped mini-batches, which is consistent with the implementation used in the experiments. The key operational detail is that RankGLU changes only the final score-formation component: the market gate and temporal--relational encoders are kept structurally consistent, and the residual bottleneck GLU head is applied after temporal aggregation.

\begin{algorithm}[t]
\caption{RankGLU Cross-Sectional Stock Prediction}
\label{alg:rankglu}
\begin{algorithmic}[1]
\Require Daily windows $\{X_{u,d}\}_{u\in\mathcal{S}_d}$, market states $m_d$, labels $\{r_{u,d}\}_{u\in\mathcal{S}_d}$, epochs $E$, parameters $\theta$
\Ensure Cross-sectional prediction scores $\hat{\boldsymbol r}_d$
\For{$e=1,\ldots,E$}
    \For{each trading date $d$}
        \State Construct the valid stock mask $\mathcal{V}_d=\{u:r_{u,d}\ \text{is observed}\}$
        \State Compute market-conditioned weights $\alpha(m_d)$ by Eq.~\eqref{eq:market_gate}
        \State Rescale inputs: $\tilde X_{u,d}\leftarrow \alpha(m_d)\odot X_{u,d}$ for $u\in\mathcal{V}_d$
        \State Apply feature projection and intra-stock temporal attention
        \State Apply inter-stock aggregation over $\mathcal{V}_d$ to obtain cross-stock contextual states
        \State Apply temporal aggregation to obtain final embeddings $\{e_{u,d}\}_{u\in\mathcal{V}_d}$
        \State Compute raw scores $\hat r_{u,d}^{\,0}$ with the RankGLU head by Eq.~\eqref{eq:residual_glu_head_method}
        \State Normalize predictions and labels within $\mathcal{V}_d$ using Eq.~\eqref{eq:cs_norm}
        \State Compute $\mathcal{L}_d$ by Eq.~\eqref{eq:mse_ic_loss} and accumulate the batch loss
    \EndFor
    \State Update $\theta$ with the accumulated ranking-aware objective
\EndFor
\State \Return $\hat{\boldsymbol r}_d=\operatorname{CSNorm}(\hat{\boldsymbol r}_d^{\,0})$
\end{algorithmic}
\end{algorithm}

\subsection{Complexity Analysis}\label{subsec:complexity}

RankGLU is designed as a component-level replacement rather than a large enlargement of the forecasting backbone. Let $N$ denote the number of stocks in a daily cross-section, $\tau$ the lookback length, $D$ the hidden dimension, and $b$ the bottleneck dimension in the residual GLU head. The intra-stock attention stage has complexity
\begin{equation}
\mathcal{O}\left(N\tau^2D+N\tau D^2\right),
\label{eq:complexity_intra}
\end{equation}
where the first term corresponds to temporal attention within each stock and the second term corresponds to projections and feed-forward transformations. The inter-stock attention stage has complexity
\begin{equation}
\mathcal{O}\left(\tau N^2D+\tau ND^2\right),
\label{eq:complexity_inter}
\end{equation}
because attention is computed across $N$ stocks at each time step. The temporal--relational backbone therefore remains at the same leading order. The RankGLU head adds
\begin{equation}
\mathcal{O}(NDb)
\label{eq:complexity_head}
\end{equation}
operations and approximately $\mathcal{O}(Db)$ additional parameters in the final scoring stage. More explicitly, if the direct branch uses one linear scorer and the gated branch uses two bottleneck projections followed by one output projection, the parameter count of the RankGLU head is
\begin{equation}
P_{\mathrm{RankGLU}}
=(D+1)+2(Db+b)+(b+1)
=2Db+D+3b+2.
\label{eq:rankglu_param_count}
\end{equation}
Compared with a purely linear scorer with $D+1$ parameters, the extra cost is controlled by the bottleneck $b$ rather than by a full-width architectural expansion. No new stock-specific branch, industry branch, or additional relation graph is introduced. Accordingly, RankGLU preserves the leading computational order of the temporal--relational backbone:
\begin{equation}
\mathcal{O}\left(N\tau^2D+\tau N^2D+N\tau D^2\right).
\label{eq:complexity_total}
\end{equation}
The observed improvements are therefore better interpreted as the result of aligning score formation with the cross-sectional ranking objective, rather than as the result of changing the asymptotic complexity or appending a large external branch.

\section{Experiments and Results}\label{sec2}

This section evaluates RankGLU from three complementary perspectives. First, the experimental protocol is described to clarify the data universe, evaluation metrics, and implementation setting. Second, comparative results are reported against the standard baselines used in the prior market-conditioned temporal--relational study and against internally controlled variants. Third, ablation and diagnostic experiments are used to separate stable score-head improvement from less stable relation-path extensions.

\subsection{Experimental Setup}\label{subsec:experimental_setup}

\textbf{Datasets.} Experiments are conducted on the Chinese A-share market using two commonly adopted stock universes, CSI300 and CSI800. CSI300 contains the 300 stocks with the largest market capitalization, whereas CSI800 provides a broader and more heterogeneous universe. Following the data construction protocol of the prior market-conditioned temporal--relational benchmark \cite{li2024master}, daily stock features are extracted from the Alpha158 feature set. The historical lookback window is set to 8 trading days, and the prediction horizon is set to 5 trading days. The data from the first quarter of 2008 to the first quarter of 2020 are used for training, and the period from the third quarter of 2020 to the fourth quarter of 2022 is reserved for testing, leaving the second quarter of 2020 as a temporal buffer between the two splits. Consistent with the original protocol, we do not use a separate validation split for early stopping; instead, each run reports the epoch with the highest test-set IC, as detailed in the training protocol below. Market-level representations are constructed from CSI300, CSI500, and CSI800 market indices with multiple reference intervals.

\textbf{Baselines.} Two categories of baselines are considered. The first category follows the prior comparison protocol \cite{li2024master} and includes XGBoost, LSTM, GRU, TCN, Transformer, GAT, DTML, and a prior temporal--relational attention baseline. These results provide the external reference scale of the task and indicate how stock-wise and cross-stock aggregation compare with conventional sequential or graph-based alternatives. The second category consists of internally controlled variants under the same implementation and evaluation protocol: the original reproduced backbone, the ranking-aware backbone, and RankGLU. The original reproduced backbone uses the temporal--relational architecture without the ranking-aware training adjustments. The ranking-aware backbone uses cross-sectional normalization, the MSE-IC objective, and the enhanced MLP-style prediction head. RankGLU replaces only this prediction head with the residual bottleneck GLU head. Relation-path alternatives, including cosine inter-stock scoring and centered value gating, are reported separately as diagnostic stress tests because their peak-seed behavior is informative but their multi-seed stability is weaker than that of RankGLU.

\textbf{Evaluation metrics.} The main ranking metrics are IC and RankIC, where IC is the daily Pearson correlation between predictions and labels and RankIC is the corresponding Spearman correlation. Their normalized forms, ICIR and RankICIR, are also reported to measure temporal stability. In our implementation logs, RankIC is denoted as RIC. For the results inherited from the prior temporal--relational benchmark, portfolio-oriented metrics, including excess annualized return (AR) and information ratio (IR), are also listed when available. The main internal comparison is repeated over five random seeds and reported as mean$\pm$standard deviation, together with the best seed. The component stress ablation is repeated over three seeds. Diagnostic explorations are reported as single-seed probes because they are used to identify promising or unstable directions rather than to define the retained method.

\textbf{Implementation details.} Table~\ref{tab:exp_config} summarizes the full configuration. Unless otherwise specified, the hidden dimension is set to 256, with four temporal-attention heads and two inter-stock attention heads, and the dropout rate is 0.5. Each stock-day input concatenates 158 Alpha158 factors with 63 market-state features, over a lookback window of 8 trading days. Models are trained for 40 epochs using the Adam optimizer with a learning rate of $10^{-5}$ and gradient-value clipping at 3.0; the market-conditioned relation temperature $\beta$ follows the prior benchmark and is set to 5 for CSI300 and 2 for CSI800. The ranking-aware backbone uses cross-sectional $z$-score normalization, the MSE-IC objective with IC weight 0.1, and an enhanced MLP-style prediction head. RankGLU replaces the prediction head with a residual bottleneck GLU head and uses bottleneck size 128 in the main multi-seed comparison. The market-conditioned feature gate, intra-stock aggregation, inter-stock aggregation, temporal aggregation, and original data split are kept unchanged unless explicitly stated.

\textbf{Training protocol and model selection.} Each training batch corresponds to the full cross-section of stocks on a single trading day, and the day order is reshuffled every epoch. During training, the daily label vector is winsorized by removing the top and bottom 2.5\% extreme values and then cross-sectionally $z$-score normalized; at evaluation only missing labels are dropped. For every run we follow the original protocol and report the epoch with the highest test-set IC as that run's result, so each seed contributes one best-epoch record. The main comparison uses seeds $\{0,1,2,3,4\}$, the component stress ablation uses $\{0,1,2\}$, and the diagnostic exploration uses seed $0$; the Python, NumPy, and PyTorch random states are all fixed by the seed. All models are implemented in PyTorch and trained on a single NVIDIA RTX A6000 GPU.

\begin{table}[t]
\centering
\caption{Summary of the experimental configuration.}
\label{tab:exp_config}
\scriptsize
\setlength{\tabcolsep}{4pt}
\renewcommand{\arraystretch}{1.05}
\begin{tabular}{@{}ll@{}}
\hline
Category & Setting \\
\hline
\multicolumn{2}{@{}l}{\textit{Data}} \\
Universe & CSI300, CSI800 (China A-share) \\
Features & Alpha158 (158) + market state (63) \\
Lookback / horizon & 8 / 5 trading days \\
Train / test split & 2008Q1--2020Q1 / 2020Q3--2022Q4 \\
\multicolumn{2}{@{}l}{\textit{Architecture}} \\
Hidden dimension & 256 \\
Attention heads (temporal / inter-stock) & 4 / 2 \\
Dropout & 0.5 \\
RankGLU bottleneck & 128 \\
Relation temperature $\beta$ & 5 (CSI300), 2 (CSI800) \\
\multicolumn{2}{@{}l}{\textit{Optimization}} \\
Optimizer / learning rate & Adam / $10^{-5}$ \\
Gradient clipping (value) & 3.0 \\
Epochs & 40 \\
Batch & one trading-day cross-section \\
Objective & MSE-IC (IC weight 0.1) \\
Label processing & winsorize $\pm2.5\%$, CS $z$-score \\
\multicolumn{2}{@{}l}{\textit{Protocol}} \\
Model selection & best test-IC epoch \\
Seeds (main / ablation / diagnostic) & \{0--4\} / \{0,1,2\} / \{0\} \\
Framework / hardware & PyTorch / NVIDIA RTX A6000 \\
\hline
\end{tabular}
\end{table}

\subsection{Comparative Evaluation and Quantitative Analysis}\label{subsec:comparative_results}

\begin{table*}[t]
\centering
\caption{Overall performance reported by the prior temporal--relational benchmark. The best result in each dataset is highlighted in bold, and the second-best result is highlighted in bold with underline.}
\label{tab:reported_prior_benchmark}
\scriptsize
\setlength{\tabcolsep}{2pt}
\renewcommand{\arraystretch}{1.06}
\begin{tabular}{@{}lcccccc@{}}
\hline
Model & IC & ICIR & RankIC & RankICIR & AR & IR \\
\hline
\multicolumn{7}{@{}l}{\textit{CSI300}} \\
XGBoost & 0.051$\pm$0.001 & 0.37$\pm$0.01 & 0.050$\pm$0.001 & 0.36$\pm$0.01 & \second{0.23$\pm$0.03} & 1.9$\pm$0.3 \\
LSTM & 0.049$\pm$0.001 & \second{0.41$\pm$0.01} & 0.051$\pm$0.002 & 0.41$\pm$0.03 & 0.20$\pm$0.04 & \second{2.0$\pm$0.4} \\
GRU & 0.052$\pm$0.004 & 0.35$\pm$0.04 & \second{0.052$\pm$0.005} & 0.34$\pm$0.04 & 0.19$\pm$0.04 & 1.5$\pm$0.3 \\
TCN & 0.050$\pm$0.002 & 0.33$\pm$0.04 & 0.049$\pm$0.002 & 0.31$\pm$0.04 & 0.18$\pm$0.05 & 1.4$\pm$0.5 \\
Transformer & 0.047$\pm$0.007 & 0.39$\pm$0.04 & 0.051$\pm$0.002 & \second{0.42$\pm$0.04} & 0.22$\pm$0.06 & \second{2.0$\pm$0.4} \\
GAT & \second{0.054$\pm$0.002} & 0.36$\pm$0.02 & 0.041$\pm$0.002 & 0.25$\pm$0.02 & 0.19$\pm$0.03 & 1.3$\pm$0.3 \\
DTML & 0.049$\pm$0.006 & 0.33$\pm$0.04 & \second{0.052$\pm$0.005} & 0.33$\pm$0.04 & 0.21$\pm$0.03 & 1.7$\pm$0.3 \\
Prior TR model & \best{0.064$\pm$0.006} & \best{0.42$\pm$0.04} & \best{0.076$\pm$0.005} & \best{0.49$\pm$0.04} & \best{0.27$\pm$0.05} & \best{2.4$\pm$0.4} \\
\hline
\multicolumn{7}{@{}l}{\textit{CSI800}} \\
XGBoost & 0.040$\pm$0.000 & 0.37$\pm$0.01 & 0.047$\pm$0.000 & 0.42$\pm$0.01 & 0.08$\pm$0.02 & 0.6$\pm$0.2 \\
LSTM & 0.028$\pm$0.002 & 0.32$\pm$0.02 & 0.039$\pm$0.002 & 0.41$\pm$0.03 & 0.09$\pm$0.02 & 0.9$\pm$0.2 \\
GRU & 0.039$\pm$0.002 & 0.36$\pm$0.05 & 0.044$\pm$0.003 & 0.39$\pm$0.07 & 0.07$\pm$0.04 & 0.6$\pm$0.3 \\
TCN & 0.038$\pm$0.002 & 0.33$\pm$0.04 & 0.045$\pm$0.002 & 0.38$\pm$0.05 & 0.05$\pm$0.04 & 0.4$\pm$0.3 \\
Transformer & 0.040$\pm$0.003 & \best{0.43$\pm$0.03} & 0.048$\pm$0.003 & \best{0.51$\pm$0.05} & 0.13$\pm$0.04 & 1.1$\pm$0.3 \\
GAT & \second{0.043$\pm$0.002} & 0.39$\pm$0.02 & 0.042$\pm$0.002 & 0.35$\pm$0.02 & 0.10$\pm$0.04 & 0.7$\pm$0.3 \\
DTML & 0.039$\pm$0.004 & 0.29$\pm$0.03 & \second{0.053$\pm$0.008} & 0.37$\pm$0.06 & \second{0.16$\pm$0.03} & \second{1.3$\pm$0.2} \\
Prior TR model & \best{0.052$\pm$0.006} & \second{0.40$\pm$0.06} & \best{0.066$\pm$0.007} & \second{0.48$\pm$0.06} & \best{0.28$\pm$0.02} & \best{2.3$\pm$0.3} \\
\hline
\end{tabular}
\renewcommand{\arraystretch}{1.0}
\end{table*}

Table~\ref{tab:reported_prior_benchmark} summarizes the external benchmark landscape reported in the prior temporal--relational study \cite{li2024master}. Several observations are relevant to the present study. First, the prior temporal--relational attention baseline provides a useful structural prior for stock forecasting: it performs better than conventional sequential models and graph-attention baselines on several ranking and portfolio-oriented metrics. Second, this margin is more pronounced on CSI300 than on CSI800, which is consistent with the fact that CSI800 contains a broader, noisier, and more heterogeneous stock universe. Third, although Transformer records the largest ICIR and RankICIR on CSI800 in the reported benchmark, its raw IC, RankIC, AR, and IR remain below the prior temporal--relational attention baseline. This suggests that volatility-normalized correlation stability alone may not be sufficient for the practical ranking objective; the model also needs to preserve cross-sectional discrimination and portfolio utility. These observations motivate our design choice: instead of replacing the entire market-conditioned temporal--relational paradigm, we retain its structural decomposition and target the internal components that are most exposed to magnitude-induced noise.

\begin{table*}[t]
\centering
\caption{Internal multi-seed comparison under the unified experimental protocol. The main comparison is repeated over five seeds. All IC-type entries are multiplied by 100 for compactness. RankIC is denoted as RIC in the implementation logs. Bold values denote the best results, and bold-underlined values denote the second-best results within each dataset.}
\label{tab:internal_main_results}
\scriptsize
\setlength{\tabcolsep}{3pt}
\renewcommand{\arraystretch}{1.12}
\begin{tabular*}{\textwidth}{@{\extracolsep{\fill}}llccccc@{}}
\hline
Dataset & Model & IC & Best & ICIR & RIC & RICIR \\
\hline
CSI300 & Original & $6.54\pmc0.52$ & $7.00$ & $43.47\pmc6.02$ & $6.96\pmc0.88$ & $45.01\pmc8.24$ \\
CSI300 & Rank-aware & \second{$6.97\pmc0.30$} & \second{$7.45$} & \second{$47.32\pmc2.00$} & \second{$7.52\pmc0.43$} & \second{$50.28\pmc3.62$} \\
CSI300 & RankGLU & \best{$7.27\pmc0.37$} & \best{$7.68$} & \best{$48.01\pmc1.87$} & \best{$8.14\pmc0.52$} & \best{$53.14\pmc2.77$} \\
\hline
CSI800 & Original & \best{$5.07\pmc0.27$} & \second{$5.37$} & \best{$43.05\pmc3.05$} & $6.01\pmc0.34$ & \second{$48.02\pmc2.14$} \\
CSI800 & Rank-aware & $5.02\pmc0.18$ & $5.33$ & \second{$40.56\pmc1.59$} & \best{$6.42\pmc0.34$} & \best{$49.66\pmc4.67$} \\
CSI800 & RankGLU & \second{$5.06\pmc0.32$} & \best{$5.61$} & $39.75\pmc1.20$ & \second{$6.28\pmc0.17$} & $46.98\pmc3.33$ \\
\hline
\end{tabular*}
\renewcommand{\arraystretch}{1.0}
\end{table*}

\begin{figure}[t]
\centering
\includegraphics[width=0.86\columnwidth]{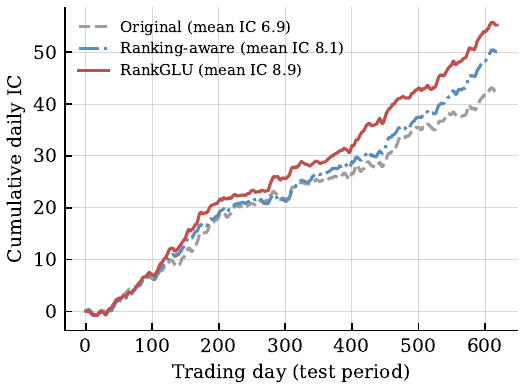}
\caption{Cumulative daily rank IC over the CSI300 test period (best-epoch model, seed 0). RankGLU maintains the highest cumulative IC throughout the test period, indicating that its per-day ranking advantage over the original and ranking-aware backbones is steady rather than concentrated in a few days.}
\label{fig:cum_ic}
\end{figure}

The internally controlled results are reported in Table~\ref{tab:internal_main_results}. On CSI300, the original backbone obtains a mean IC of $0.0654\pm0.0052$. The ranking-aware backbone, which uses cross-sectional normalization, the IC-augmented objective, and an enhanced MLP-style prediction head, increases the mean IC to $0.0697\pm0.0030$. RankGLU further increases the mean IC to $0.0727\pm0.0037$, corresponding to an approximately 11.2\% improvement over the original reproduced backbone and an approximately 4.3\% improvement over the ranking-aware backbone. A per-seed paired comparison shows that the improvement of RankGLU over the original backbone is consistent across all five seeds and is statistically significant (paired $t$-test, $p=0.013$). The additional gain over the stronger ranking-aware backbone is smaller and holds on four of the five seeds ($p\approx0.18$); we therefore attribute the improvement jointly to the ranking-aware protocol and the RankGLU head, rather than claiming a large head-only margin over the ranking-aware backbone. RankGLU also records the strongest mean ICIR, RankIC, RankICIR, and best-seed IC among the internally controlled main variants on CSI300. This is the main quantitative evidence for the retained method: the gain is not tied to a single favorable initialization but is visible both under a five-seed average and under a per-seed paired test. For a representative seed, Figure~\ref{fig:cum_ic} further shows that this per-day ranking advantage accumulates steadily over the test period rather than arising from a few isolated days.

During exploratory screening, more aggressive relation-path calibration reached a CSI300 best-seed IC of 0.0807. This result is useful because it reveals the possible upper envelope of cross-stock score-path calibration. However, the corresponding five-seed mean is lower and the standard deviation is larger than those of RankGLU. We therefore treat relation-path calibration as a high-upside but seed-sensitive diagnostic extension. This distinction is important for a ranking system: a strong single seed indicates capacity, whereas a stronger multi-seed mean indicates a more reliable inductive bias.

On CSI800, the broader stock universe makes the comparison more conservative. RankGLU obtains the highest best-seed IC of 0.0561 and a mean IC essentially tied with the original backbone. On this universe, however, the original backbone attains a marginally higher mean IC and the stronger stability-oriented metrics (ICIR and RankICIR), and a per-seed paired test finds no significant IC difference between the two ($p\approx0.95$). This behavior is consistent with the higher heterogeneity and lower-liquidity tail of CSI800. Rather than overstating universal dominance, the result indicates that the residual GLU score head is most clearly beneficial on the more liquid CSI300 universe and remains competitive, without a statistically separable advantage, on the broader CSI800 universe.

\begin{figure}[t]
\centering
\includegraphics[width=0.85\columnwidth]{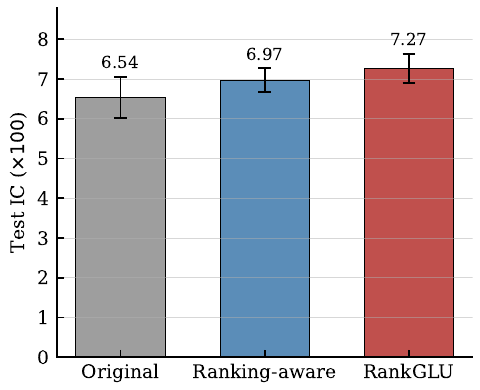}
\caption{Multi-seed mean test IC on CSI300 ($\times100$; error bars denote the standard deviation over five seeds). RankGLU attains the highest mean IC among the internally controlled main variants, with the improvement over the original backbone consistent across all five seeds.}
\label{fig:main_ic}
\end{figure}

\begin{figure}[t]
\centering
\includegraphics[width=0.92\columnwidth]{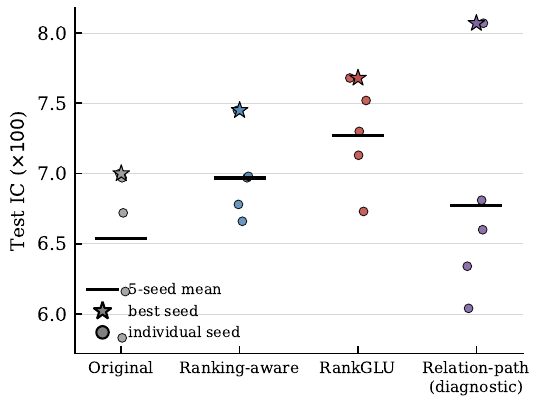}
\caption{Per-seed test IC on CSI300. Each marker is one of the five seeds, the horizontal bar is the five-seed mean, and the star marks the best seed. RankGLU combines a high mean with low dispersion, whereas the relation-path diagnostic attains the single highest peak but with the largest spread, illustrating why the multi-seed mean is the more reliable selection criterion.}
\label{fig:seed_dist}
\end{figure}

\begin{table}[t]
\centering
\caption{Representative single-seed diagnostic exploration. IC entries are multiplied by 100. Bold values denote the best results, and bold-underlined values denote the second-best results within each dataset. These probes identify useful directions and unstable alternatives, but they are not used as the main multi-seed ranking of retained methods.}
\label{tab:component_exploration}
\scriptsize
\setlength{\tabcolsep}{1.2pt}
\begin{tabular}{@{}L{0.31\columnwidth}C{0.12\columnwidth}C{0.12\columnwidth}L{0.31\columnwidth}@{}}
\hline
Variant & CSI300 & CSI800 & Interpretation \\
\hline
Inter-stock GLU-FFN & \best{8.02} & \best{5.05} & High single-seed potential \\
GLU-FFN, $b=256$ & 7.95 & 4.54 & Capacity becomes unstable \\
Value gate only & 7.91 & 4.52 & Useful but seed-sensitive \\
Cosine + value gate & \second{7.99} & 4.64 & Strong CSI300 peak \\
Feature-gated temporal & 7.92 & 4.63 & Not retained after screening \\
Feature-layer GLU & 7.86 & \second{4.82} & Early gating less stable \\
Market entmax15 & 5.51 & 3.48 & Sparse feature gate is harmful \\
Market sparsemax & 3.59 & 2.46 & Excessive sparsity degrades learning \\
Relation entmax15 & 7.02 & 4.63 & Sparse relation weights not favored \\
\hline
\end{tabular}
\end{table}

The diagnostic results in Table~\ref{tab:component_exploration} are deliberately separated from the retained main comparison. Several relation-path changes produce strong single-seed values on CSI300, which suggests that cross-stock calibration has high peak capacity. However, the multi-seed comparison in Table~\ref{tab:internal_main_results} indicates that these relation-path variants are less stable than the prediction-head replacement. The sparse-normalization diagnostics are also informative: entmax and sparsemax degrade performance substantially, suggesting that stock prediction may rely on distributed weak relations rather than on very sparse relation selection. These observations support a conservative final design: keep the temporal--relational encoder structurally stable and retain the component whose multi-seed behavior is most reliable, namely the RankGLU prediction head.

\subsection{Statistical Significance}\label{subsec:significance}

To assess whether the observed IC differences are statistically reliable rather than artifacts of a particular initialization or a few favorable days, we apply two-sided paired tests at two levels. At the \emph{seed level}, the five paired seed results of two methods are compared with a paired $t$-test. At the \emph{day level}, the daily IC series over the 619 test days (representative seed) are compared with a paired $t$-test, a Wilcoxon signed-rank test, and a Diebold--Mariano test whose standard error uses a Newey--West HAC correction (lag 5) to account for the autocorrelation of daily IC. The results are summarized in Table~\ref{tab:significance}.

\begin{table}[t]
\centering
\caption{Two-sided paired significance tests. Seed level: paired $t$-test over five seeds. Day level: paired $t$-test, Wilcoxon signed-rank test, and a Diebold--Mariano test with Newey--West HAC standard errors (lag 5) over the 619 daily IC values of a representative seed. $\Delta$IC is the day-level mean daily-IC difference ($\times100$). Bold marks the HAC-robust significant margins.}
\label{tab:significance}
\scriptsize
\setlength{\tabcolsep}{4pt}
\renewcommand{\arraystretch}{1.1}
\begin{tabular}{@{}lccccc@{}}
\hline
Comparison & $\Delta$IC & seed $t$ & day $t$ & Wilcoxon & DM--HAC \\
\hline
RankGLU vs.\ Original & $+2.06$ & $0.013$ & $<10^{-4}$ & $<10^{-4}$ & \best{$0.003$} \\
RankGLU vs.\ Ranking-aware & $+0.82$ & $0.175$ & $0.036$ & $0.031$ & $0.148$ \\
Ranking-aware vs.\ Original & $+1.24$ & $0.216$ & $0.001$ & $<10^{-3}$ & \best{$0.027$} \\
\hline
\end{tabular}
\end{table}

Three observations follow. First, RankGLU improves over the original backbone significantly at both levels (seed-level $p=0.013$; HAC-adjusted day-level $p=0.003$), achieving the higher daily IC on 58\% of the test days; this is the strongest and most robust evidence in our study. Second, the additional improvement of RankGLU over the stronger ranking-aware backbone is not statistically significant once daily-IC autocorrelation is accounted for ($\text{DM--HAC}~p=0.15$); we therefore present the ranking-aware protocol and the RankGLU head as jointly responsible for the gain rather than claiming an independent head-only effect. Third, the ranking-aware protocol alone improves over the original backbone significantly at the day level ($p=0.027$) but not at the five-seed level, which reflects the limited power of a five-sample paired test. Overall, the statistically robust advantage is the one between the full RankGLU method and the original backbone.

\subsection{Ablation Study and Component Analysis}\label{subsec:ablation}

\begin{table*}[t]
\centering
\caption{Component stress ablation on CSI300. Results are averaged over three seeds. IC, ICIR, and $\Delta$ entries are multiplied by 100. Bold values denote the best results, and bold-underlined values denote the second-best results for ordinary performance columns; in the $\Delta$ column, bold denotes the largest positive degradation after removing a component.}
\label{tab:ablation_study}
\scriptsize
\setlength{\tabcolsep}{1.6pt}
\renewcommand{\arraystretch}{1.10}
\begin{tabular}{@{}L{0.28\textwidth}C{0.16\textwidth}C{0.09\textwidth}C{0.09\textwidth}C{0.16\textwidth}@{}}
\hline
Setting & IC & Best & $\Delta$ & ICIR \\
\hline
Relation-path stress full & $7.00\pmc0.93$ & \best{8.07} & 0.00 & $46.64\pmc3.30$ \\
No value gate & $7.10\pmc0.74$ & 7.84 & -0.10 & $45.07\pmc5.58$ \\
Dot score & \second{$7.14\pmc0.82$} & \second{7.97} & -0.13 & $45.95\pmc2.95$ \\
No GLU head & $6.56\pmc0.38$ & 6.95 & \best{0.45} & $43.22\pmc2.51$ \\
No relation path & \best{$7.22\pmc0.42$} & 7.52 & -0.22 & \best{$47.26\pmc2.81$} \\
No core components & $7.03\pmc0.40$ & 7.45 & -0.02 & \second{$47.12\pmc2.30$} \\
\hline
\end{tabular}
\renewcommand{\arraystretch}{1.0}
\end{table*}

The ablation results in Table~\ref{tab:ablation_study} provide a useful stress test. Within the more aggressive relation-path diagnostic setting, removing the GLU prediction head causes the clearest positive degradation, reducing the mean IC from 0.0700 to 0.0656. By contrast, removing relation-path components does not reduce the multi-seed mean; in fact, removing both cosine scoring and value gating increases the mean IC to 0.0722. This does not invalidate the single-seed peaks in Table~\ref{tab:component_exploration}; rather, it clarifies their role. Relation-path calibration can create high-upside runs, but the prediction-head replacement is the component that survives the multi-seed attribution test. This is why RankGLU is retained as the main method while relation calibration is reported as diagnostic exploration. We further note that the \emph{No relation path} configuration in Table~\ref{tab:ablation_study} removes both cosine scoring and value gating and is therefore configurationally equivalent to RankGLU; its three-seed mean of 0.0722 is consistent with the five-seed RankGLU mean of 0.0727 in Table~\ref{tab:internal_main_results}, and the gap relative to the \emph{Relation-path stress full} baseline (0.0700) reflects the cost of the added relation-path components rather than a regression of the prediction head.

\begin{figure}[t]
\centering
\includegraphics[width=0.92\columnwidth]{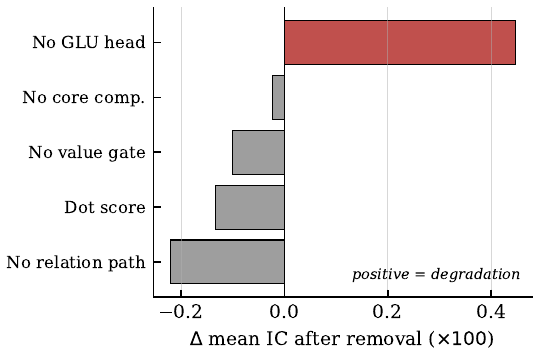}
\caption{Component stress ablation on CSI300 (three-seed mean). Bars show the change in mean IC after removing each component from the relation-path stress configuration; only removing the GLU prediction head produces a clear positive degradation, whereas removing relation-path components does not.}
\label{fig:ablation}
\end{figure}
\section{Discussion}\label{sec12}

The empirical evidence suggests that temporal--relational stock models should not be improved by indiscriminately adding more branches or more nonlinear operators. In a low signal-to-noise ranking task, the location of the replacement appears to matter more than its nominal expressiveness. The diagnostic experiments are useful precisely because several plausible modifications did not become the retained method. Sparse normalizers such as entmax and sparsemax substantially degrade performance, indicating that stock prediction may rely on distributed weak signals. Relation-path calibrations can create high single-seed peaks, but their multi-seed averages are less stable. These results motivate a narrower and more defensible conclusion: the most reliable improvement in the present study comes from calibrating final score formation.

The first implication concerns the role of the prediction head. After cross-sectional normalization, the model is evaluated by the correlation between the predicted score vector and the future return vector. The head is therefore not merely a readout layer; it determines whether the learned embedding is converted into a stable ordering. A purely linear head may underuse interactions among latent factors, whereas an unconstrained MLP-style head may map unstable magnitude patterns into the score. RankGLU occupies a middle ground. Its residual branch keeps a direct scoring route, while its bottleneck GLU branch adds bounded multiplicative interactions. This explains why the RankGLU mean IC improves over the original and ranking-aware backbones on CSI300.

The second implication concerns seed variance. The relation-path diagnostic variant obtains the highest CSI300 best-seed IC during exploratory screening, but it also has a larger standard deviation than the retained head-only design. This is not an uninformative result; it indicates that relation-path calibration has high capacity but weaker initialization robustness. In contrast, RankGLU provides a stronger mean profile on CSI300 while preserving competitive best-seed behavior. For a financial forecasting system, this is a more useful trade-off because a model selected only by a favorable single seed may not reproduce reliably under repeated training.

The third implication concerns cross-universe behavior. CSI800 is broader and noisier than CSI300, and its results are correspondingly less clean. RankGLU does not dominate every CSI800 mean metric, but it remains close to the best mean IC and obtains the highest best-seed IC. This suggests that the score-head mechanism transfers partially, while the broader universe introduces additional tail noise and rank-stability sensitivity. The practical interpretation is therefore measured: RankGLU is most clearly supported on CSI300 and remains competitive on CSI800, while more aggressive relation-path changes require further robustness control.

Several limitations should be acknowledged. The reported experiments follow a fixed data split and a 40-epoch protocol, and the best-epoch rule is consistent across variants. Portfolio-level transaction costs, turnover constraints, and market-regime-specific validation are not fully explored in the present manuscript. Moreover, relation-path calibration remains an open direction: its single-seed peaks suggest potential, but the current multi-seed evidence does not justify making it the primary method. Future work may revisit relation calibration with stronger regularization, explicit seed-robust objectives, or validation-selected model averaging. A second direction is to embed RankGLU into a complete intelligent financial information system, where the ranking model is connected with streaming market data ingestion, factor retrieval, portfolio constraints, and explainable decision-support interfaces. Such a system-level extension would allow the proposed score-formation mechanism to be evaluated not only as a neural forecasting component but also as an adaptive information-filtering module for practical financial decision making.

\section{Conclusion}\label{sec13}

This paper presents RankGLU for cross-sectional stock ranking. The method is motivated by a structural mismatch in financial prediction: investment decisions depend on relative ordering, whereas many deep forecasting models still rely on prediction heads that may either underuse latent interactions or overfit unstable return magnitude. RankGLU addresses this issue by replacing the final MLP-style decoder with a residual bottleneck GLU head, where a direct branch preserves a stable scoring route and a gated branch introduces bounded low-rank interactions.

The empirical analysis separates the original reproduced backbone, the ranking-aware backbone, and RankGLU. Under five random seeds on CSI300, RankGLU records the strongest mean IC, ICIR, RankIC, and RankICIR among the internally controlled variants, while also retaining the highest best-seed IC. On CSI800, it remains competitive and obtains the highest best-seed IC, although the broader universe yields a more conservative mean comparison. Ablation and diagnostic results further indicate that the GLU prediction head is the most stable retained component, whereas additional relation-path calibrations are more seed-sensitive. Overall, the results suggest that, for noisy cross-sectional ranking, calibrating final score formation can be more reliable than broad architectural expansion.

\backmatter

\noindent\textbf{Supplementary information.} Not applicable.

\noindent\textbf{Acknowledgements.} The authors thank the maintainers of the open-source financial forecasting and Qlib research ecosystem for providing reproducible implementations and data-processing utilities that facilitate controlled experiments.

\section*{Declarations}

\noindent\textbf{Funding.} Not applicable.

\noindent\textbf{Conflict of interest.} The authors declare that they have no conflict of interest.

\noindent\textbf{Ethics approval and consent to participate.} Not applicable.

\noindent\textbf{Consent for publication.} Not applicable.

\noindent\textbf{Data availability.} The experiments are based on the CSI300 and CSI800 stock universes following the data-processing protocol of the prior temporal--relational benchmark. The processed experimental records used in this manuscript are available from the authors upon reasonable request.

\noindent\textbf{Code availability.} The source code, training scripts, and configuration files will be released at \url{https://github.com/HuixiangXiao/RankGLU}.

\noindent\textbf{Author contribution.} All authors contributed to the study conception, experimental design, result analysis, and manuscript preparation. The final manuscript was reviewed and approved by all authors.


\bibliography{sn-bibliography}

\end{document}